\documentclass[prx,reprint,aps,amsmath,amssymb,amsfonts,superscriptaddress,notitlepage,floatfix]{revtex4-2}
\usepackage{pifont}
\usepackage[colorlinks=true,citecolor=blue,urlcolor=blue,linkcolor=blue,hyperfigures=true]{hyperref}
\usepackage{graphicx}
\usepackage{amsmath,amsfonts,amssymb}
\usepackage[table,x11names,dvipsnames]{xcolor}
\usepackage{color}
\usepackage{changes}
\usepackage{booktabs}
\usepackage{ulem}

\usepackage{xspace}
\usepackage{tikz}
\usetikzlibrary{decorations.pathmorphing}

\newcommand{\iwn}{i\omega_n}

\newcommand{\iwm}{i\nu_m}

\newcommand{\kvec}{\mathbf{k}}
\newcommand{\kpvec}{\mathbf{k}'}

\newcommand{\qvec}{\mathbf{q}}
\newcommand{\identitymatrix}{\hat{\mathcal{I}}}

\def\G0W0{G$_0$W$_0$\xspace}
\def\STO{SrTiO$_3$\xspace}
\def\MoS2{MoS$_2$\xspace}
\def\MoSe2{MoSe$_2$\xspace}
\def\WS2{WS$_2$\xspace}
\def\WSe2{WSe$_2$\xspace}

\begin{document}

\preprint{APS/123-QED}

\title{Enhancing Plasmonic Superconductivity in Layered Materials via Dynamical Coulomb Engineering}

\author{Y. in 't Veld}
\affiliation{Institute for Molecules and Materials, Radboud University, 6525 AJ Nijmegen, the Netherlands}
\author{M. I. Katsnelson}
\affiliation{Institute for Molecules and Materials, Radboud University, 6525 AJ Nijmegen, the Netherlands}
\affiliation{Constructor Knowledge Institute, Constructor University, 28759 Bremen, Germany}
\author{A. J. Millis}
\affiliation{Center for Computational Quantum Physics, Flatiron Institute, New York, NY 10010, United States of America}
\affiliation{Department of Physics, Columbia University, New York, NY 10027, United States of America}
\author{M. R\"osner}
\affiliation{Institute for Molecules and Materials, Radboud University, 6525 AJ Nijmegen, the Netherlands}

\date{\today}
\begin{abstract}

Conventional Coulomb engineering, through controlled manipulation of the environment, offers an effective route to tune the correlation properties of atomically thin van der Waals materials via static screening.
Here we present tunable \emph{dynamical} screening as a method for precisely tailoring bosonic modes to optimize many-body properties.
We show that ``bosonic engineering'' of plasmon modes can be used to enhance plasmon-induced superconducting critical temperatures of layered superconductors in metallic environments by up to an order of magnitude, due to the formation of interlayer hybridized plasmon modes with enhanced superconducting pairing strength.
We determine optimal properties of the screening environment to maximize critical temperatures. We show how bosonic engineering can aid the search for experimental verification of plasmon mediated superconductivity.
\end{abstract}

\maketitle

\textbf{Introduction.} ``Coulomb engineering" \cite{steinke_coulomb-engineered_2020,van_loon_coulomb_2023} aims to tailor many-body effects via modifications to the fundamental electron-electron interactions.
Atomically thin van-der-Waals materials are promising candidates to exploit this, because the low dimensionality enhances correlation effects while the large surface-to-volume ratio allows for efficient external manipulation. 
As a result, when van-der-Waals materials are placed on substrates or when they are encapsulated by other materials, the modified environmental screening properties can
modify the range \cite{waldecker_rigid_2019,tebbe_tailoring_2023}, strength \cite{van_loon_coulomb_2023} and dynamical behavior \cite{steinke_coulomb-engineered_2020} of the Coulomb interaction.

The environmental modification of screened dynamical interactions presents an intriguing many-body engineering route characterized by a tailored interplay of internal and external bosonic modes. 
Examples include  plasmons or phonons in the environment coupling to the Fermi sea of a layered van-der-Waals metals leaving behind additional spectral features~\cite{ulstrup_observation_2024} or substrate phonons acting on excitons in 2D semiconductors \cite{steinhoff_dynamical_2020,steinhoff_frequency-dependent_2018}.
Depending on the degree of coupling, these modes can either act individually or in a composite manner to affect the electronic properties of the two-dimensional (2D) material, opening up the possibility for ``bosonic engineering'': i.e. the control of dynamical interactions by external means. Substrate phonon enhancement of electron pairing has been proposed for superconducting monolayers~\cite{zhou_dipolar_2017,wang_phonon_2017,zhang_enhanced_2019,zhang_role_2016,aperis_multiband_2018,fatemi_synthesizing_2018,rosenstein_high-temperature_2019,wei_high-temperature_2021} and hybridized composite bosons such as plasmonic multicomponent modes or interacting plasmon-phonon modes were discussed in particular with respect to their effects on superconductivity in single-layer~\cite{in_t_veld_screening_2023} as well as stacked~\cite{kresin_layer_1988,pashitskii_plasmon_2008,grankin_interplay_2023,bill_electronic_2003} and/or twisted layered materials~\cite{guerci_topological_2024,cea_coulomb_2021,peng_theoretical_2024}.

Here we take these ideas one step further, showing how bosonic engineering via dynamical external screening of the Coulomb interaction can drastically enhance plasmon-induced superconducting critical temperatures in layered materials by strengthening the plasmon-induced superconducting pairing at low frequencies, without also increasing the high-frequency Coulomb-induced repulsion.

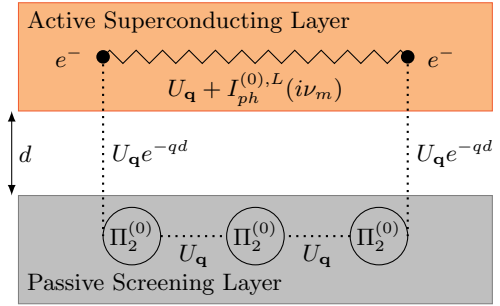
\begin{figure}[t]
    \centering
    \pgfmathsetmacro{\matWidth}{7}
\pgfmathsetmacro{\matHeight}{1.60}
\pgfmathsetmacro{\matDist}{1.25}

\pgfmathsetmacro{\activeX}{0}
\pgfmathsetmacro{\activeY}{\matHeight+\matDist}

\pgfmathsetmacro{\passiveX}{0}
\pgfmathsetmacro{\passiveY}{0}

\pgfmathsetmacro{\elecX}{1.25}
\pgfmathsetmacro{\elecY}{\activeY+\matHeight/2}

\pgfmathsetmacro{\bubbleY}{\passiveY+\matHeight/2+0.20}
\pgfmathsetmacro{\bubbleWidth}{0.85}

\begin{tikzpicture}[scale=0.9]
    \draw[black,fill=Apricot,draw=RedOrange] (\activeX,\activeY,0) -- ++(\matWidth,0,0) -- ++(0,\matHeight,0) -- ++(-\matWidth,0,0) -- cycle;
    \draw[black,fill=lightgray,draw=gray] (\passiveX,\passiveY,0) -- ++(\matWidth,0,0) -- ++(0,\matHeight,0) -- ++(-\matWidth,0,0) -- cycle;

    \node[anchor=south west] at (\passiveX,\passiveY) {Passive Screening Layer};
    
    \node[anchor=north west] at (\activeX,\activeY+\matHeight){Active Superconducting Layer};

    \draw[>=latex, <->,black] (\passiveX-0.1,\passiveY+\matHeight) -- (\activeX-0.1,\activeY);
    \node[] at (\passiveX+0.1,\passiveY+\matHeight+\matDist/2) {$d$};

    \fill[black] (\activeX+\elecX,\elecY) circle (0.1);
    \fill[black] (\activeX+\matWidth-\elecX,\elecY) circle (0.1);

    \node[] at (\activeX+\elecX-0.5,\elecY) {$e^-$};
    \node[] at (\activeX+\matWidth-\elecX+0.5,\elecY) {$e^-$};

    \draw[decorate, decoration={zigzag}] (\activeX+\elecX,\elecY) -- (\activeX+\matWidth-\elecX,\elecY);
    \node[anchor=north, font=\small] at (\activeX+\matWidth/2,\elecY-0.1) {$U_{\mathbf{q}} + I_{ph}^{(0),L}(i\nu_m)$};

    \draw[dotted,thick] (\activeX+\elecX,\elecY) -- (\activeX+\elecX,\bubbleY);
    \draw[dotted,thick] (\activeX+\matWidth-\elecX,\elecY) -- (\activeX+\matWidth-\elecX,\bubbleY);

    \node[anchor=west, font=\small] at (\activeX+\elecX,\activeX+\matHeight+\matDist/2) {$U_{\mathbf{q}} e^{-qd}$};
    \node[anchor=west, font=\small] at (\activeX+\matWidth-\elecX,\activeX+\matHeight+\matDist/2) {$U_{\mathbf{q}} e^{-qd}$};

    \draw[black] (\activeX+\elecX+\bubbleWidth/2,\bubbleY) circle (\bubbleWidth/2);
    \draw[black] (\passiveX+\matWidth-\elecX-\bubbleWidth/2,\bubbleY) circle (\bubbleWidth/2);
    \draw[black] (\passiveX+\matWidth/2,\bubbleY) circle   (\bubbleWidth/2);

    \node[font=\small] at (\activeX+\elecX+\bubbleWidth/2,\bubbleY) {$\Pi_2^{(0)}$};
    \node[font=\small] at (\passiveX+\matWidth-\elecX-\bubbleWidth/2,\bubbleY) {$\Pi_2^{(0)}$};
    \node[font=\small] at (\passiveX+\matWidth/2,\bubbleY) {$\Pi_2^{(0)}$};

    \draw[dotted,thick] (\activeX+\elecX+\bubbleWidth,\bubbleY) -- (\passiveX+\matWidth/2-\bubbleWidth/2,\bubbleY);
    \draw[dotted,thick] (\passiveX+\matWidth-\elecX-\bubbleWidth,\bubbleY) -- (\passiveX+\matWidth/2+\bubbleWidth/2,\bubbleY);

    \node[anchor=north, font=\small] at (\elecX/2 + \bubbleWidth/4 + \passiveX + \matWidth/4,\bubbleY) {$U_{\mathbf{q}}$};
    \node[anchor=north, font=\small] at (\passiveX - \bubbleWidth/4 - \elecX/2 + 3*\matWidth/4,\bubbleY) {$U_{\mathbf{q}}$};

\end{tikzpicture}
    \caption{\label{fig:systemSchematic} Schematic illustration of the heterostructure that we study. The top orange layer is the active layer, which hosts superconductivity. The bottom gray layer is a screening layer which dynamically screens the active layer via the long-range Coulomb interaction. The dotted lines represent (interlayer) bare Coulomb interactions, the circles represent polarization processes within the screening layer and the jagged line represents the bare (longitudinal) interactions within the active layer.
}
\end{figure}

\textbf{Theoretical Description.} We study the hetero-bilayer system shown schematically in Fig.~\ref{fig:systemSchematic}: an active superconducting layer, coupled via long-range Coulomb interactions to a passive metallic screening layer. We assume that the layers are electronically decoupled with no hybridization between the electronic wave functions in the two layers. This setup can for example be realized by two van-der-Waals materials separated by an insulating spacer, such as a monolayer of hexagonal Boron Nitride (hBN). 

We describe the bilayer with the following Hamiltonian
\begin{align}
    H &= \sum_{\kvec i \sigma} \varepsilon_{\kvec,i} c^{\sigma,\dagger}_{\kvec i} c^{\sigma\phantom{,\dagger}}_{\kvec i} + \frac{1}{2} \sum_{\kvec \kpvec \qvec} \sum_{ij\sigma\sigma'} \hat{U}_{ij}(\qvec) c^{\sigma,\dagger}_{\kvec+\qvec i} c^{\sigma',\dagger}_{\kpvec-\qvec j} c^{\sigma'\phantom{,\dagger}}_{\kpvec j} c^{\sigma\phantom{,\dagger}}_{\kvec i} \nonumber\\
    &+ \sum_{\qvec\nu} \omega_e b^\dagger_{\qvec \nu} b^{\phantom{\dagger}}_{\qvec \nu}  
    + \sum_{\kvec \qvec \nu \sigma} g \left( b^{\phantom{\dagger}}_{\qvec \nu} +  b^\dagger_{-\qvec \nu}\right) c^{\sigma,\dagger}_{\kvec+\qvec 1} c^{\sigma\phantom{,\dagger}}_{\kvec 1}
\end{align}
where $c^{\sigma,(\dagger)}_{\kvec i \sigma}$ and $b^{(\dagger)}_{\qvec \nu}$ are the usual electronic and phononic annihilation (creation) operators for layer $i$ and phonon mode $\nu$, respectively. We assume that each layer has a single band $\varepsilon_{\kvec,i}$ of free electrons and a Coulomb interaction $U_{\qvec}$. The active layer ($i=1$) furthermore hosts longitudinal optical (LO) and transversal optical (TO) phonon modes, with electron-phonon coupling constants $g$ and phonon energies $\omega_{e}$ which are assumed to be momentum-independent for simplicity.
The passive layer ($i=2$, assumed metallic) hosts a plasmon mode that is coupled to density fluctuations in the active layer via the long-range Coulomb interaction. 
In this situation the phonons and Coulomb interaction combine to yield respective effective LO and TO interactions $I^{(0),LO}_{ph}(\iwm)$ and $I^{(0),TO}_{ph}(\iwm)$ between the electrons in the same layer, which are accompanied by non-local Coulomb interactions within and between both layers. 
In this setup we can integrate out the electronic degrees of freedom of the passive layer, leaving behind a longitudinal interaction of the form
\begin{align}
    \tilde{I}^{(0),L}_1(\qvec,\iwm) &= U_\qvec + I^{(0),LO}_{ph}(\iwm)\nonumber\\
    &+ U_\qvec e^{-qd} \frac{\Pi^{(0)}_2(\qvec,\iwm)}{1 - U_\qvec \Pi^{(0)}_2(\qvec,\iwm)} U_\qvec e^{-qd}
    \label{eq:effBareI}
\end{align}
with $U_{\qvec} = 2 \pi e^2 / (A \varepsilon q)$, $I^{(0),LO/TO}_{ph}(\iwm)=2\omega_e g^2 / ((\iwm)^2 - \omega_e^2)$ and the polarization in the passive layer $\Pi^{(0)}_2(\qvec,\iwm)$, $A$ is the unit-cell area, $d$ the distance between the two layers and $\varepsilon$ is a global dielectric constant.
In Fig.~\ref{fig:systemSchematic} we show schematically the different terms of $\tilde{I}^{(0),L}_1(\qvec,\iwm)$ within the layer from which they originate.
The fully screened interaction is finally given by $I_1(\qvec,\iwm)=I^{L}_1(\qvec,\iwm)+I^{(0),TO}_{ph}(\iwm)$, with $I^{L}_1(\qvec,\iwm)$ obtained from screening $\tilde{I}^{(0),L}_1(\qvec,\iwm)$ within the random phase approximation using the polarization in the superconducting layer $\Pi^{(0)}_1(\qvec,\iwm)$. Further details including the derivation and detailed specification of the model can be found in the supplemental information (SI).

In the high-frequency limit $I_1(\qvec,\iwm \rightarrow \infty) = U_\qvec$ is simply the bare Coulomb repulsion in the superconducting layer, which is unaffected by screening from the passive layer. The passive screening layer thus tunes only the low-energy part of the interaction, without altering the high-energy repulsive part.

Bosonic engineering occurs here via tuning material properties of the passive layer and the distance between the passive and active layers to optimize superconductivity in the active layer.
We use this model within a fully retarded and non-local one-loop Eliashberg solver as applied in Ref.~\cite{in_t_veld_screening_2023}, to investigate how the distance $d$ as well as the electronic properties (effective mass and the doping level) of the passive metallic layer control the superconducting properties of the active layer. 

\begin{figure}[t!]
    \centering
    \includegraphics[width=0.99\linewidth]{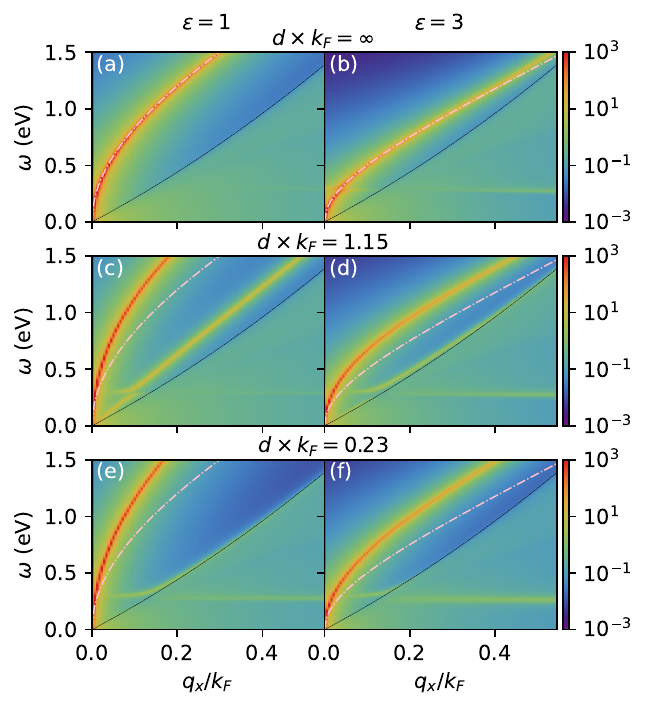}
    \caption{\label{fig:plasmonsDispersion} The effective interaction $2 \alpha^2 F^L_\qvec(\omega) / \omega$ within the active layer, for interlayer distances $d$ shown in the legend at $\varepsilon=1$ (left panels) and $\varepsilon=3$ (right panels). The dashed black lines denote the upper edges of the electron-hole continua and the dashed pink lines denote the plasmon dispersion of an isolated layer (i.e., at $d \times k_F = \infty$). 
    These results were obtained at $T = 100$\,K, with $m^* = 0.2 m_e$ and $E_{F} = 1$\,eV in both layers.
}
\end{figure}

\textbf{Tunable Composite Bosonic Modes.} 
We start by investigating how bosonic Coulomb engineering affects $\alpha^2F^L_\qvec(\omega) = -N_0 \text{Im} I^L_1(\qvec,\omega) / \pi$, the spectral function characterizing the longitudinal pairing modes in the active layer.  
Fig.~\ref{fig:plasmonsDispersion} depicts $2 \alpha^2 F^L_\qvec(\omega) / \omega$ for various interlayer distances $d$ and environmental dielectric constants $\varepsilon$. 
For infinite layer distance [panels (a) and (b)] the passive layer is irrelevant and we reproduce the hybridized phonon-plasmon mode of a monolayer in a static dielectric environment \cite{in_t_veld_screening_2023}. In this case, at $\varepsilon = 3$ there is hybridization between the dispersionless phonon mode and the $\sqrt{q}$-like two-dimensional plasmon mode, whereas at $\varepsilon = 1$ the Coulomb interaction screens out the phonon-mode, such that the phonon-plasmon hybridization is negligible. As the interlayer distance is decreased to $d\times k_F = 1.15$ at $\varepsilon = 1$ [panel (c)], two distinct inter-layer plasmon modes appear, as is typical of coupled two-dimensional plasmon modes~\cite{das_sarma_collective_1981,santoro_acoustic_1988,hwang_plasmon_2009,sensarma_dynamic_2010}.
The high-energy mode can be understood as an in-phase oscillation of the total charge in both layers, which still has a $\sqrt{q}$-like dispersion in the long-wavelength limit, but shifted to higher energies compared to the plasmon mode at $d = \infty$ (indicated by the dashed pink line). 
The low-energy mode can be described as an out-of-phase dipolar oscillation between the charge densities in the two layers.
It has a linear dispersion and lies below the energy of the isolated monolayer plasmon mode. As the interlayer distance decreases further to $d \times k_F = 0.23$ [panel (e)], the out-of-phase mode shifts into the electron-hole continuum (boundary indicated by the dashed black lines), such that it is Landau damped if the layers are close enough. Also the LO phonon appears as an evident absorption feature at larger momenta.

Similar trends can be observed for $\varepsilon = 3$ [panels (d) and (f)], but in this case the distance at which the out-of-phase mode starts to be Landau damped is larger.
As for the longitudinal phonon mode, its coupling strength to the electrons (reflected by its intensity in Fig.~\ref{fig:plasmonsDispersion}) is not significantly affected by the dynamical interlayer screening. This is a consequence of the different energy scales of the phonon and plasmon modes. It does, however, hybridize more strongly with the dipolar plasmon mode than with the charged plasmon mode.

This shows that the individual bosons from the passive and active layers interact and hybridize in conceptually simple but quantitatively non-trivial ways, forming composite excitations with tunable dispersions and coupling strengths.

\begin{figure}
    \centering
    \includegraphics[width=0.99\linewidth]{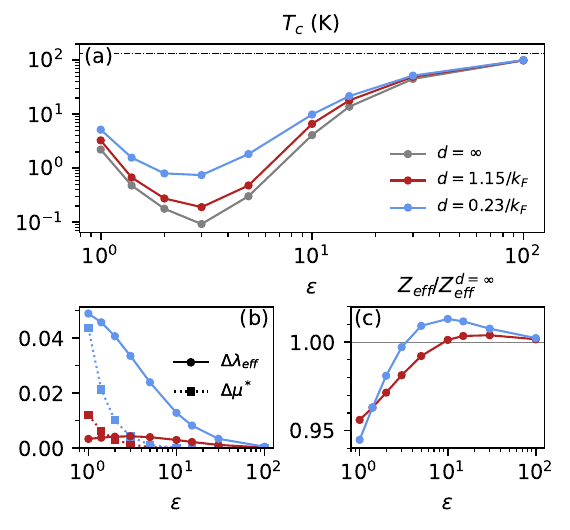}
    \caption{\label{fig:SCState} (a) The superconducting critical temperature $T_c$ within the active layer as a function of external screening $\varepsilon$, for different interlayer distances $d$. The horizontal dashed line denotes the $T_c$ corresponding to $\varepsilon = \infty$. (b) The change of the effective pairing strength $\Delta \lambda_{\text{eff}} = \lambda_{\text{eff}} - \lambda_{\text{eff}}^{d=\infty}$ (solid lines) and of the TMA pseudo-potential $\Delta \mu^* = \mu^* - \mu^*_{d=\infty}$ (dotted lines) upon introducing a metallic screening layer. (c) The relative change of the effective mass-renormalization factor $Z_{\text{eff}} / Z_{\text{eff}}^{d=\infty}$. 
    These results were obtained at $T = 100$\,K, with $m^* = 0.2 m_e$ and $E_{F} = 1$\,eV in both layers.
    }
\end{figure}

\textbf{Superconductivity.} Fig.~\ref{fig:SCState}(a) shows the superconducting critical temperatures $T_c$ for different interlayer distances $d$ as well as for different overall Coulomb interaction strengths tuned by $\varepsilon$.
The infinite separation trace (gray line) reproduces previous results \cite{in_t_veld_screening_2023}.
At $\varepsilon = \infty$ the only interactions contributing to superconductivity are the local LO and TO electron-phonon interactions and correspondingly we find negligible distance dependence of the critical temperature. As $\varepsilon$ is decreased from $\infty$ to $\sim 20$ the calculated transition temperature decreases markedly because the high frequency Coulomb repulsion $U_\qvec$ increases as $\propto 1/\varepsilon$, but the dynamical Coulomb interaction remains weak and thus the distance dependence of the transition temperature remains small. In this ``phononic'' regime environmental bosonic engineering is ineffective.
As $\varepsilon$ decreases further, $T_c$ reaches a minimum around $\varepsilon \approx 3$ while the dependence of the transition temperature on the interlayer distance becomes much stronger. Here the dominant effect is the continuing dramatic increase of the high frequency instantaneous Coulomb repulsion. However, the plasmonic pairing also begins to play a role, and may be engineered by the presence of the passive layer.
Beyond this minimum, for $\varepsilon \lesssim 2$, the plasmon mediated electron-electron attraction starts to dominate the interaction, leading again to an enhancement of $T_c$. In this ``plasmonic'' regime the effect of the passive layer remains large.

To qualitatively disentangle the different contributions to the superconducting state, we use standard formulae (see SI) to approximate the full frequency dependent interaction in terms of an effective electron-electron coupling $\lambda_{\text{eff}}$, an effective boson frequency $\omega_{\text{eff}}$, a Tolmachev-Morel-Anderson (TMA) Coulomb pseudo-potential $\mu^*$ and a mass-renormalization factor $Z_{\text{eff}}$ as used in the conventional discussions in phonon-mediated superconductors. With these parameters, we can calculate an effective critical temperature that qualitatively reproduces the trends of our numerical results, but is quantitatively off. As in discussions of conventional superconductivity, this analysis should thus be understood as a tool to gain intuitive understanding but cannot be used for quantitative predictions.

In Fig.~\ref{fig:SCState}(b) we show that $\lambda_{\text{eff}}$ increases with decreasing distance $d$ as well as with decreasing $\varepsilon$. Thus, proximity to the screening layer enables a stronger coupling of external dynamical boson modes to the electrons in the active superconducting layer. This is clearly a supporting mechanism of bosonic Coulomb engineering.
However, we also see that $\mu^*$ increases with decreasing $d$ and decreasing $\varepsilon$, which is counteracting the $\lambda_{\text{eff}}$ enhancement.
As argued before, the bare Coulomb potential is not affected by the interlayer coupling. However, the TMA pseudo-potential $\mu^*$ can be altered by interlayer coupling via changes in the effective boson frequency $\omega_{\text{eff}}$, which suggests an additional direction for bosonic engineering.
This effect is mostly negligible in the phononic regime, but in the plasmonic regime $\varepsilon \lesssim 3$ it causes a significant enhancement of $\mu^*$, as a consequence of the enhanced $\omega_{\text{eff}}$ (see Supplement).
Nevertheless, for most values of $\varepsilon$, $\lambda_{\text{eff}}$ is enhanced more than $\mu^*$. 
In Fig.~\ref{fig:SCState}(c) we further show the relative change of $Z_{\text{eff}}$. Interestingly, the mass-renormalization is enhanced in the phononic regime, whereas it is reduced in the plasmonic regime.
The actual changes are, however, only on the order of a few percent, such that mass-renormalization alone cannot explain the $T_c$ enhancement.
We therefore conclude that the driving force behind the enhancement of $T_c$ is the enhancement of the effective electron-electron attraction $\lambda_{\text{eff}}$, which is, however, weakened through an enhanced $\mu^*$ in the low $\varepsilon$ regime as a consequence of the enhanced $\omega_{\text{eff}}$.
Our results thus show that bosonic Coulomb engineering via dynamical external screening can be utilized to strongly enhance plasmonic superconductivity in layered materials, leading to critical temperatures increased by factors of up to $20$ as discussed in the following.

\textbf{Optimizing Superconductivity.} 
To enhance $T_c$ the screening layer should lead to a low-energy composite bosonic mode, which couples strongly to the electrons. 
This could be achieved by tuning the electronic properties of the screening layer. In our model this is controlled by its effective mass as well as by its doping level. While the carrier concentration in the passive layer can be controlled by gating, the effective mass is difficult to control in situ. However, passive layers could be chosen to have heavy mass carriers.

\begin{figure}[t!]
    \centering
    \includegraphics[width=0.99\linewidth]{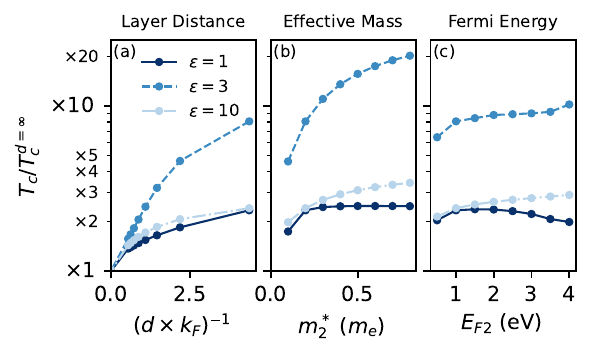}
    \caption{\label{fig:optimizingTc} The enhancement of the superconducting critical temperature with respect to an isolated monolayer $T_c / T^{d=\infty}_c$, as a function of the inverse interlayer distance (a), the effective mass of the screening layer (b) and the Fermi energy of the screening layer (c). We use as a baseline $d \times k_F = 0.23$, $m^*_2 = 0.2 m_e$ and $E_{F2} = 1$\,eV and vary the respective parameters from there.
    }
\end{figure}

In Fig.~\ref{fig:optimizingTc} we show the relative change of the full numerically evaluated $T_c$ due to the presence of the passive layer, as a function of different tuning parameters. 
In panel (a) we again find a significant $T_c$ enhancement as the interlayer distance is reduced. Especially at $\varepsilon = 3$ this enhancement is a factor of 8 for the smallest $d$ considered, but also for $\varepsilon = 1$ and $\varepsilon = 10$ we find up to a doubling of the critical temperature.
In Fig.~\ref{fig:optimizingTc}(b) we show the results when tuning the effective mass of the passive layer $m_2^*$ at fixed interlayer distance $d\times k_F = 0.23$. In these calculations the total electron density was kept fixed by adjusting the Fermi energy $E_{F2}$ correspondingly.
As we show in the supplement, enhancing $m_2^*$ pushes all plasmon modes as well as the edge of the electron-hole continuum of the screening layer to lower energies.
As a consequence, $\lambda_{\text{eff}}$ is enhanced and $\mu^*$ is reduced as $m_2^*$ is increased. This interplay of enhancing $\lambda_{\text{eff}}$ and reducing $\mu^*$ is optimal to increase $T_c$, leading to the strong enhancement in Fig.~\ref{fig:optimizingTc}(b).
Overall we conclude that a metallic screening layer with a large effective mass is most favorable for enhancing plasmonic pairing in a layered superconductor.

In Fig.~\ref{fig:optimizingTc}(c) we summarize the effect of tuning the Fermi energy of the screening layer $E_{F2}$ at fixed effective mass $m_2^*$, thereby changing the electron density in the screening layer.
Enhancing $E_{F2}$ causes the in-phase plasmon mode to shift to higher energies, as shown in the supplement. This leads to an enhancement of $\mu^*$, while $\lambda_{\text{eff}}$ is relatively unaffected, such that the combined $\Delta \lambda_{\text{eff}} - \Delta \mu^*$ suggests a reduction of the critical temperature with increased doping.
Since $Z_{\text{eff}}$ is reduced as well, $T_c$ is ultimately relatively unaffected by modifications to the doping in the screening layer.
Nevertheless, we see a strong, but not doping-tunable $T_c$ enhancement as a result of the presence of the screening layer.

\textbf{Conclusions.} 
We have shown that in all cases considered, bosonic engineering of plasmonic modes via interlayer dynamical screening can significantly enhance the superconducting critical temperature of two-dimensional materials.
Especially in the regime of intermediate Coulomb interaction strengths, where in the monolayer limit $T_c$ has a minimum due to the unfavorable interplay of static Coulomb repulsion and dynamic pairing, $T_c$ can be enhanced by more than an order of magnitude.
The driving mechanism behind this enhancement is the formation of composite interlayer plasmon modes, which have a total coupling strength to the electrons that is larger than that of the individual monolayer plasmon modes.
The enhanced coupling strength is however counteracted by changes in the Coulomb pseudopotential $\mu^*$ and the mass-renormalization $Z_{\text{eff}}$, which are also affected by dynamical interlayer screening.
Our results show that, in order to get the most favorable combination of all these competing effects for high $T_c$, the electrons in the screening layer should have large effective mass. 
We furthermore find that the electron density of the screening layer is significantly less relevant than the effective mass for engineering optimal $T_c$.

Beyond the implications for bosonic engineering, our work shows that appropriately designed structures can provide evidence relevant to the long-standing question of plasmon mediated (or enhanced) superconductivity.  The  balance between static repulsion and dynamical attraction defines $T_c$ and one would like to know to which extent dynamical screening associated with plasmons can lead to  pairing. We find that proximity to a passive layer can enhance superconductivity in the plasmonic regime, but not in the phononic regime; thus observation of increases in $T_c$ due to the presence of a passive layer indicates that plasmonic pairing is in play.
To this end, the electron-doped semiconducting transition metal dichalcogenides (TMDCs) are especially promising. 
Some of the TMDCs, such as \MoS2\cite{lu_evidence_2015} and \WS2\cite{zheliuk_monolayer_2017}, have been shown to superconduct in the monolayer limit upon electron doping. Moreover, the normal state of TMDCs is known to be sensitive to (dynamical) environmental screening \cite{waldecker_rigid_2019,ulstrup_observation_2024} and the conduction band minimum is well described by an effective mass approximation \cite{haastrup_computational_2018,gjerding_recent_2021}. Heterostructures of such a superconducting TMDC monolayer and another metallic monolayer with larger effective mass, such as NbTe$_2$ \cite{haastrup_computational_2018,gjerding_recent_2021}, separated by a hBN monolayer might therefore be an experimental realization of the model discussed here. 

Our results could furthermore be relevant in the observed $T_c$ enhancement in electron-doped semi-conducting TMDC multi-layers as the number of layers is increased \cite{costanzo_gate-induced_2016,khestanova_unusual_2018,benyamini_fragility_2019}.
Previous work has shown that effects from static screening alone cannot explain this behavior \cite{schonhoff_interplay_2016}, but additional pairing strength from interlayer plasmon modes might explain the $T_c$ enhancement.

Finally, bosonic engineering is not restricted to plasmon modes alone. 
Other bosons in the environment might similarly form interlayer composite boson modes whenever their dispersions overlap with boson dispersions in the active layer and whenever their excitation energies lie outside the regime of Landau damping. 
This might be relevant for the drastic $T_c$ enhancement in monolayer FeSe on \STO, which has been argued to be induced by \STO phonon modes coupled into the FeSe layer \cite{lee_interfacial_2014,wang_interface-induced_2012}.
From our results we hypothesize that this enhancement may instead result from the renormalization of FeSe boson modes (e.g., phonons, plasmons, or magnons) due to hybridization with external \STO bosonic modes, and not from pairing induced by \STO phonons on their own.

\acknowledgements{AJM acknowledges support from the Keele foundation in the context of the Super-C collaboration. MIK acknowledges support from the European Research Council through
the ERC Synergy Grant 854843-FASTCORR. MR acknowledges support from the Vidi ENW research programme of the Dutch Research Council (NWO) under the grant https://doi.org/10.61686/YDRHT18202 with file number VI.Vidi.233.077. YV and MR acknowledge support from the Dutch Research Council (NWO) via the ``TOPCORE'' consortium.}

\onecolumngrid
\newpage
\section{Supplemental Information}

\subsection{Model}
We study a hetero-bilayer system consisting of two layers which are electronically decoupled. The electronic structure of both layers is described by an effective-mass approximation $\varepsilon_{\kvec,i} = k^2 / (2 m_i^*) - E_{F,i}$, with $m_i^*$ and $E_{F,i}$ the effective mass and Fermi energy in layer $i$, respectively, where $i=1$ is the active superconducting layer and $i = 2$ is the passive screening layer. We consider the following Hamiltonian
\begin{align}
    H &= \sum_{\kvec i \sigma} \varepsilon_{\kvec,i} c^{\sigma,\dagger}_{\kvec i} c^{\sigma\phantom{,\dagger}}_{\kvec i} + \frac{1}{2} \sum_{\kvec \kpvec \qvec} \sum_{ij\sigma\sigma'} \hat{U}_{ij}(\qvec) c^{\sigma,\dagger}_{\kvec+\qvec i} c^{\sigma',\dagger}_{\kpvec-\qvec j} c^{\sigma'\phantom{,\dagger}}_{\kpvec j} c^{\sigma\phantom{,\dagger}}_{\kvec i} \nonumber\\
    &+ \sum_{\qvec\nu} \omega_e b^\dagger_{\qvec \nu} b^{\phantom{\dagger}}_{\qvec \nu}  
    + \sum_{\kvec \qvec \nu \sigma} g \left( b^{\phantom{\dagger}}_{\qvec \nu} +  b^\dagger_{-\qvec \nu}\right) c^{\sigma,\dagger}_{\kvec+\qvec 1} c^{\sigma\phantom{,\dagger}}_{\kvec 1},
\end{align}
where $c^{\sigma,(\dagger)}_{\kvec i}$ is the annihilation (creation) operator of electrons in layer $i$ with momentum $\kvec$ and spin $\sigma$. $b^{(\dagger)}_{\qvec \nu}$ is the annihilation (creation) operator of phonons with transfer momentum $\qvec$, which we include in the active superconducting layer only. Here $\nu \in \{\text{LO},\text{TO}\}$ for the longitudinal optical (LO) and transverse optical (TO) phonon mode, respectively, $g$ is the electron-phonon coupling constant and $\omega_{e}$ is the phonon energy, which we both assume to be local for simplicity.
The matrix elements of the long-range Coulomb interaction are given in the layer-basis by
\begin{align}
    \hat{U}(\qvec)  &= U_{\qvec}
    \begin{pmatrix} 
        1  &  e^{-qd} \\
        e^{-qd}   &  1
    \end{pmatrix},
\end{align}
where $U_{\qvec} = 2 \pi e^2 / (A \varepsilon q)$, with $A$ the unit-cell area, $d$ the distance between the two layers and $\varepsilon$ a global dielectric constant.
In this basis the diagonal and off-diagonal components represent intra- and interlayer interactions, respectively.
The total screened interaction matrix in the RPA is given by
\begin{align}
    \hat{I}(\qvec,\iwm) &= \left[ \identitymatrix - \left(\hat{U}(\qvec) + \hat{I}^{(0),LO}_{ph}(\qvec,\iwm)\right) \hat{\Pi}^{(0)}(\qvec,\iwm) \right]^{-1} \left(\hat{U}(\qvec) + \hat{I}^{(0),LO}_{ph}(\qvec,\iwm)\right) \nonumber\\
    &+ \hat{I}^{(0),TO}_{ph}(\iwm),
\end{align}
where 
\begin{align}
    \hat{I}^{(0),LO/TO}_{ph}(\qvec,\iwm)  &= 
    \begin{pmatrix} 
        I^{(0),TO/LO}_{ph}(\iwm) &  0 \\
        0   &  0
    \end{pmatrix},
\end{align}
$I^{(0),LO/TO}_{ph}(\iwm)=2\omega_e g^2 / ((\iwm)^2 - \omega_e^2)$ and the RPA polarization matrix $\hat{\Pi}^{(0)}(\qvec,\iwm)$ is diagonal as a consequence of the vanishing interlayer electronic hybridization. The TO phonon mode is assumed to be unscreened in the RPA. 

From here, we apply a downfolding procedure to integrate out the passive layer, leaving behind an effective single-layer model with a renormalized screened interaction given by $I_1(\qvec,\iwm) = I^L_1(\qvec,\iwm) + I^{(0),TO}_{ph}(\iwm)$. 
The longitudinal part $I^L_1(\qvec,\iwm)$ is given by
\begin{equation}
    I^L_1(\qvec,\iwm) = \frac{\tilde{I}^{(0),L}_1(\qvec,\iwm)}{1 - \tilde{I}^{(0),L}_1(\qvec,\iwm) \Pi^{(0)}_1(\qvec,\iwm)},
\end{equation}
with the effective interaction $\tilde{I}^{(0),L}_1(\qvec,\iwm)$ given in the main text and $\Pi^{(0)}_1(\qvec,\iwm)$ the RPA polarization in the active superconducting layer.
We fix for the active layer $m_1^* = 0.2\,m_e$ and $E_{F1} = 1$\,eV, while tuning $m_2^*$ and $E_{F2}$ of the passive layer. Here $m_e$ is the free electron mass. For both the LO and TO bare phonon modes we set $\omega_e = 0.3$\,eV and $g^2 = 0.3$\,eV$^2$.

\subsection{Computational Details}
All calculations were performed using the TRIQS\cite{parcollet_triqs_2015} and TPRF\cite{nils_wentzell_triqstprf_2024} codebases, using a linearly discretized momentum mesh of 800x800 points. The Matsubara axis was represented using the recently developed discrete Lehman representation (DLR)\cite{kaye_discrete_2022}, which drastically reduces the temperature scaling of the required amount of Matsubara frequencies to $\mathcal{O}(\text{log}(\beta))$ compared to a full Matsubara mesh which scales as $\mathcal{O}(\beta)$. Due to the improved scaling with temperature, we can resolve the superconducting critical temperatures in the low screening limit. For all calculations, the DLR error tolerance was set to $\epsilon = 10^{-10}$ and the high-energy cutoff to $\omega_{\text{max}} = 50$\,eV. Critical temperatures were obtained by evaluating the leading eigenvalue $\lambda(T)$ of the linearized gap equation at logarithmically spaced temperatures between $1$\,K and $100$\,K. We then perform a linear fit of $\lambda(T)$ as a function of $\text{log}(T)$, which yields $T_c$ as the temperature at which $\lambda(T) = 1$.
Real-frequency \G0W0 calculations (not shown in the main text) have been performed on a 100x100 momentum mesh and a linearly spaced frequency mesh with 1000 points between -20 and 20 eV.

\subsection{Normal-State}

\begin{figure*}[h]
    \centering
    \includegraphics[scale=0.7]{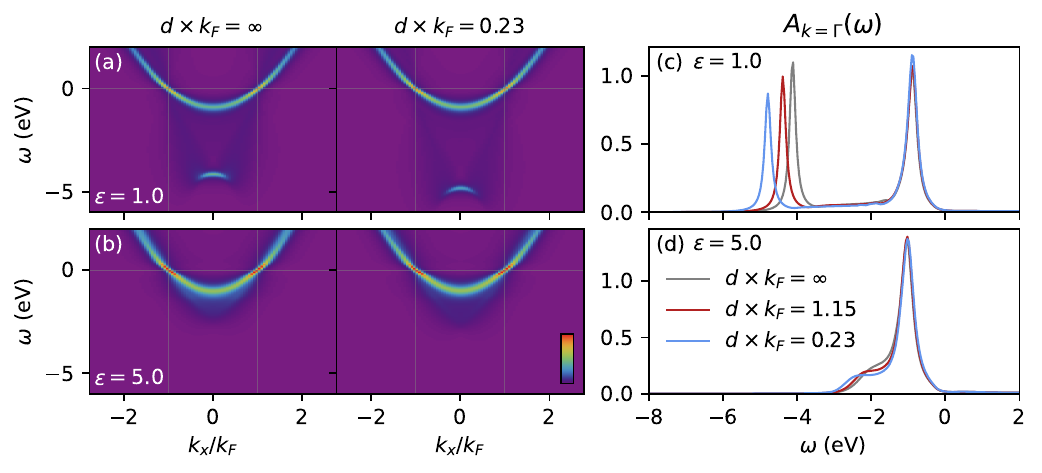}
    \caption{\label{fig:normalState} (a,b) Momentum-resolved \G0W0 spectral function $A_\kvec(\omega)$, for different interlayer distances at $\varepsilon=1$ (a) and $\varepsilon=5$ (b). (c,d) Linecut of the spectral function $A_\kvec(\omega)$ at $\kvec = \Gamma$ at $\varepsilon=1$ and $\varepsilon=5$, respectively. These results were obtained for equivalent monolayers, with $m_1^* = m_2^* = 0.2 m_e$ and $E_{F1} = E_{F2} = 1$\,eV. The temperature is set to $T = 100$\,K.}
\end{figure*}

In the left panels of Fig.~\ref{fig:normalState} (a) and (b) we show the dressed electronic spectral function $A_\kvec(\omega) = -\text{Im} G(\kvec,\omega) / \pi$ in the \G0W0 approximation of a monolayer in a static dielectric environment. These results are reminiscent to those obtained in Ref.~\cite{in_t_veld_screening_2023} for a square lattice. Around the Fermi energy at $\varepsilon = 5$ we find the typical phononic mass-enhancement, which is strongly reduced for $\varepsilon = 1$ due to the screening from the Coulomb interaction. Below the band minimum we find additional spectral weight coming from plasmon polaron excitations. For $\varepsilon = 1$ these excitations induce a relatively coherent shakeoff band, whereas for $\varepsilon = 5$ they induce an incoherent shoulder. The energy separation between the shakeoff feature and the band minimum is determined by a representative energy scale of the plasmon dispersion, as discussed in Ref.~\cite{ulstrup_observation_2024}.
When we introduce the passive screening layer [shown in the right panels of Fig.~\ref{fig:normalState}(a) and (b)], we find qualitatively the same features. Quantitatively, however, the plasmon polaron shakeoff features shift to lower energies, further away from the band minimum, as seen in the frequency linecuts of $A_{\kvec}(\omega)$ at $\kvec = \Gamma$ [shown in panels (c) and (d) of Fig.~\ref{fig:normalState}]. This is a consequence of the shift of the plasmonic spectral weight to higher energies, due to the enhanced frequency of the charged interlayer plasmon mode and due to the Landau damping of the neutral interlayer plasmon mode.

\begin{figure*}[t]
    \centering
    \includegraphics[scale=0.7]{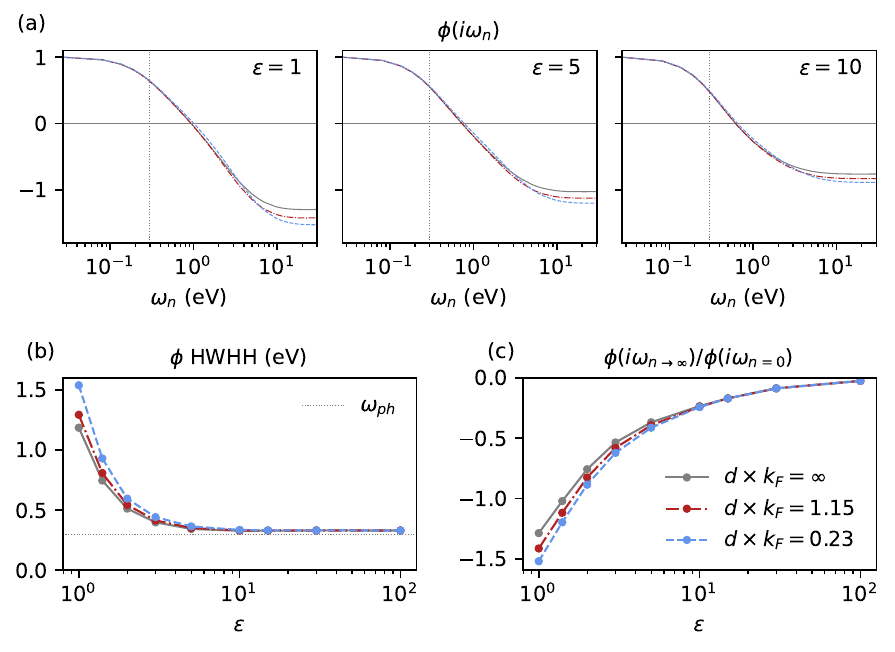}
    \caption{\label{fig:SCStatePhi} (a) The momentum-summed anomalous self-energy $\phi(i\omega_n)$ at $T = 100$\,K as a function of Matsubara frequency $\omega_n$. The left, middle and right panels have $\varepsilon=1$, $5$ and $10$, respectively. The vertical dotted line indicates the bare phonon frequency $\omega_e$. In all cases $\phi$ is normalized such that $\phi(i\omega_{n=0}) = 1$. (b) The half-width at half-height of $\phi(i\omega_n)$. (c) The high-frequency tail of $\phi(i\omega_n)$. These results were obtained for equivalent monolayers, with $m_1^* = m_2^* = 0.2 m_e$ and $E_{F1} = E_{F2} = 1$\,eV.
}
\end{figure*}

\subsection{Anomalous Self-Energy}
In Fig.~\ref{fig:SCStatePhi}(a) we show the local (momentum-summed) anomalous self-energy $\phi(\iwn) = \sum_{\kvec} \phi(\kvec,\iwn)$. Since $\phi(\kvec,\iwn)$ is obtained from the linearized gap equation, it has an arbitrary normalization factor. Here, the normalization factor was fixed such that $\phi(i\omega_{n=0}) = 1$. For all $d$ we find the characteristic features of the crossover from phonon to plasmon mediated superconductivity as $\varepsilon$ is reduced, as discussed in Ref.~\cite{in_t_veld_screening_2023}.
In short, the increased half-width at half-height (HWHH) of $\phi(\iwn)$ [shown in panel (b)] as $\varepsilon$ reduces reflects that the dominating mediating boson switches from the lower-energy phonon mode to the higher-energy plasmon mode. 
Furthermore, the high-frequency tail [shown in panel (c)] is reduced as $\varepsilon$ reduces due to the enhanced high-frequency Coulomb repulsion at low $\varepsilon$.

Introducing the neighboring metallic layer does not qualitatively change the trends of $\phi(\iwn)$ as a function of $\varepsilon$, indicating that there is still a crossover from phonon to plasmon mediated superconductivity. 
Quantitatively there are some differences however. Firstly, the half-width at half-height of $\phi(\iwn)$ is enhanced by interlayer dynamical screening. Similarly to the changes in the normal state spectral function, this reflects that the spectral weight of the interaction is shifted to higher energies.
Secondly, the high-frequency tail of $\phi(\iwn)$ is reduced upon reducing $d$.
This can be analyzed in an approximate BCS picture, in which the high-frequency tail of $\phi(\iwn)$ behaves as $-\mu^* / (\lambda_{\text{eff}} - \mu^*)$, with $\mu^*$ the Tolmachev-Morel-Anderson (TMA) Coulomb pseudopotential and $\lambda_{\text{eff}}$ being an effective parameter describing the electron-electron attraction.\cite{morel_calculation_1962,coleman_introduction_2015}
From the qualitative modeling in the main text we expect the value of $\lambda_{\text{eff}}$ to increase as the layers are brought closer together. The reduced high-frequency tail therefore indicates that the pseudopotential $\mu^*$ is also enhanced upon reduced $d$.
These results agree with our qualitative modeling and again hint towards a delicate balance between dynamical attraction, and (renormalized) static repulsion.

\subsection{Qualitative Modeling}
Our qualitative modeling of the different contributions to the superconducting state closely follows the McMillan-Allen-Dynes description for phonon-mediated superconductors.\cite{carbotte_properties_1990,mcmillan_transition_1968,allen_transition_1975,dynes_mcmillans_1972}
We define the following expression for the effective critical temperature
\begin{equation}
    k_B T_c^{\text{eff}} = 1.13 \omega_{\text{eff}} \text{exp} \left( -\frac{Z_{\text{eff}}}{\lambda_{\text{eff}} - \mu^*} \right).
\end{equation}
Here the effective dimensionless electron-electron pairing strength $\lambda_{\text{eff}}$ is defined as the momentum integral of the zero-frequency electron-electron coupling
\begin{align}
    \lambda_{\text{eff}} &= 
    - N_0 \sum_\qvec \left( I_1(\qvec,i\nu_{m=0}) - U_\qvec \right) \nonumber\\
    &= \int_0^\infty d\omega \sum_\qvec \frac{2 \alpha^2 F^L_\qvec(\omega)}{\omega}
    + \frac{2 g^2 N_0}{\omega_e}.
\end{align}
This effective coupling is counteracted by the static Coulomb repulsion. Similar to the coupling, we will define a corresponding dimensionless parameter $\mu^C$. 
A common definition of $\mu^C$ is a double Fermi surface average of $U_{\qvec}$.\cite{morel_calculation_1962,bogoljubov_new_1958,simonato_revised_2023} Here, we instead define it by the momentum sum $\mu^C = N_0 \sum_\qvec U_\qvec$ for consistency with $\lambda_{\text{eff}}$.
From Eliashberg theory we understand, however, that it is not the bare potential $\mu^C$ that enters the effective low-energy gap-equation, but the renormalized TMA pseudo-potential $\mu^*$\cite{tolmachev_logarithmic_1961,morel_calculation_1962}, given by
\begin{equation}
    \mu^* = \frac{\mu^C}{1 + \mu^C \text{log}\left(E_B / \omega_{\text{eff}}\right)},
\end{equation}
where $E_B$ is the bandwidth (set here to $E_B = 4 E_{F1}$ for simplicity). The effective boson frequency $\omega_{\text{eff}}$ we define using the logarithmic average
\begin{align}
    \omega_{\text{eff}} &= \text{exp}\left[ \frac{2}{\lambda_{\text{eff}}} \int_0^\infty d\omega \sum_\qvec \alpha^2 F^L_\qvec(\omega) \frac{\text{log}(\omega)}{\omega} \right] \nonumber\\
    &\times \text{exp} \left[ \frac{2}{\lambda_{\text{eff}}} \frac{g^2 N_0 \text{log}(\omega_e)}{\omega_e} \right].
\end{align}
The final contribution to the critical temperature is the mass-renormalization factor $Z_{\text{eff}}$. Similar to Eliashberg theory, we define it using the frequency derivative of the \G0W0 self-energy evaluated at $k_F$
\begin{equation}
    Z_{\text{eff}} = 1 - \left.\frac{\partial \Sigma(k_F,\iwn)}{\partial (\iwn)} \right\vert_{\iwn=0}.
\end{equation}

\subsection{Optimizing Superconductivity}

\begin{figure}[t!]
    \centering
    \includegraphics[scale=1.0]{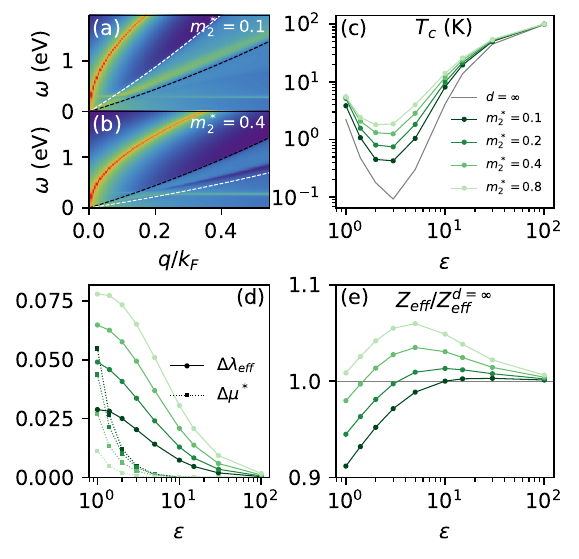}
    \caption{\label{fig:varyEffMass} Variations in the effective mass of the screening layer $m_2^*$ at fixed density. The interlayer distance is set to $d \times k_F = 0.23$ and $T = 100$\,K. $m_2^*$ is given in units of $m_e$ in the figure. (a,b) The factor $2 \alpha^2 F^L_\qvec(\omega) / \omega$ at $\varepsilon=1$ for two effective masses $m_2^*$ of the screening layer. The black and white dashed lines are the upper edges of the electron-hole continua of the superconducting and screening layers, respectively. (c) The superconducting critical temperature as a function of $\varepsilon$, for various $m_2^*$. (d) The change of the effective pairing strength (solid lines) and effective TMA pseudopotential (dotted lines) with respect to the isolated monolayer. (e) The relative change of the mass-renormalization factor with respect to the isolated monolayer.
    }
\end{figure}

In Fig.~\ref{fig:varyEffMass} we summarize the results when tuning the effective mass of the screening layer $m_2^*$ at fixed interlayer distance $d\times k_F = 0.23$. In these calculations the total electron density was kept fixed by changing the Fermi energy $E_{F2}$ correspondingly.
In panels (a) and (b) we show $2 \alpha^2 F^L_\qvec(\omega) / \omega$ in the superconducting layer, for two different screening layer effective masses. 
Enhancing $m_2^*$ shifts both hybridized plasmon modes to lower energies. For $m_2^* = 0.1\,m_e$ [panel (a)] the out-of-phase dipolar mode is still above the continuum of the superconducting layer, but is broadened by the continuum of the screening layer. As $m_2^*$ is increased to $0.4\,m_e$ [panel (b)] it gets damped further, such that it has negligible spectral weight and thus a vanishing coupling to the electrons.
The in-phase plasmon mode, has a large $(\qvec,\omega)$ space where it is undamped. Notably, for $m_2^* = 0.1\,m_e$ this region is limited by the continuum of the screening layer (white dashed line), whereas for $m_2^* = 0.4\,m_e$ it is limited by the continuum of the superconducting layer (black dashed line). Therefore, besides shifting down the plasmon dispersions, enhancing $m_2^*$ also increases the spectral weight of the in-phase plasmon mode by shifting away the electron-hole continuum of the screening layer. This effect is reflected in the enhancement of $\lambda_{\text{eff}}$ shown in panel (d). In contrast, the Coulomb pseudopotential $\mu^*$ is reduced for enhanced $m_2^*$. This is a consequence of the reduced energy of the in-phase plasmon mode. This interplay of enhancing $\lambda_{\text{eff}}$ and reducing $\mu^*$ is optimal to increase $T_c$, as we see in Fig.~\ref{fig:varyEffMass}~(c).
The relative $T_c$ enhancement is, however, not the same for all $\varepsilon$, with at $\varepsilon = 1$ significantly smaller changes than around $\varepsilon = 3$.
This results from the larger mass-renormalization factor $Z_{\text{eff}}$ shown in panel (e).

\begin{figure*}[t!]
    \centering
    \includegraphics[scale=1.0]{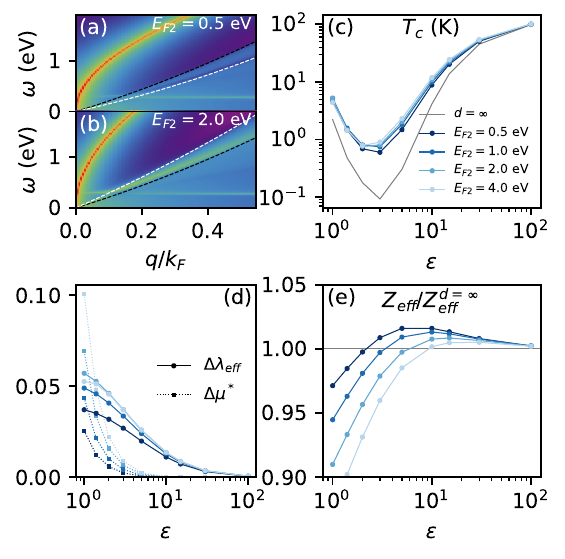}
    \caption{\label{fig:varyEf} Variations in the Fermi energy of the screening layer $E_{F2}$. The interlayer distance is set to $d \times k_F = 0.23$ and $T = 100$\,K. (a,b) The factor $2 \alpha^2 F^L_\qvec(\omega) / \omega$ for $\varepsilon=1$ and $\varepsilon=5$, respectively, for two Fermi energies. The black and white dashed lines are the upper edges of the electron-hole continua of the active layer and screening layers, respectively. (c) The superconducting critical temperature as a function of $\varepsilon$, for various $E_{F2}$. (d) The change of the effective pairing strength (solid lines) and effective TMA pseudopotential (dotted lines) with respect to the isolated monolayer. (e) The relative change of the mass-renormalization factor with respect to the isolated monolayer.
}
\end{figure*}

In Fig.~\ref{fig:varyEf} we summarize the effect of tuning the Fermi energy of the neighboring layer $E_{F2}$ at fixed effective mass $m_2^*$, thereby changing the electron density in the screening layer.
In panels (a) and (b) we show $2 \alpha^2 F^L_\qvec(\omega) / \omega$ in the superconducting layer, for two different screening layer Fermi energies. 
Enhancing $E_{F2}$ causes the in-phase plasmon mode to shift to higher energies. As shown in panel (d), this leads to an enhancement of $\mu^*$, while $\lambda_{\text{eff}}$ is relatively unaffected, such that the combined $\Delta \lambda_{\text{eff}} - \Delta \mu^*$ suggests a reduction of the critical temperature with increased doping.
However, since $Z_{\text{eff}}$ [panel (e)] is reduced as well, $T_c$ [panel (c)] is relatively unaffected by modifications to the doping in the screening layer.

\bibliography{bib}

\end{document}